 \documentclass[default]{sn-jnl}
\usepackage{graphicx}%
\usepackage{multirow}%
\usepackage{amsmath,amssymb,amsfonts}%
\usepackage{amsthm}%
\usepackage{mathrsfs}%
\usepackage[title]{appendix}%
\usepackage{xcolor}%
\usepackage{textcomp}%
\usepackage{manyfoot}%
\usepackage{booktabs}%
\usepackage{algorithm}%
\usepackage{algorithmicx}%
\usepackage{algpseudocode}%
\usepackage{listings}%
\usepackage{tabularx}  % For more flexible tables
\usepackage{adjustbox}
\usepackage{graphicx}   % For including graphics
\usepackage{subcaption}
\usepackage{multicol}
\usepackage{multirow}
\usepackage[utf8]{inputenc}
\usepackage{subcaption}
   \usepackage{natbib}
  \usepackage{amsfonts}
  \usepackage[T1]{fontenc}    % Font encoding
 \usepackage{anysize}
 \usepackage{setspace}
\usepackage{tfrupee}
\newtheorem{definition}{Definition}

\newtheorem{theorem}{Theorem}%  meant for continuous numbers
\usepackage{lmodern}
\usepackage{anyfontsize}
\begin{document}

\title[Article Title]{ Analysis of the impact of the union budget announcements on the Indian stock market: A fractal perspective}

% Fractal dimensional analysis of the union budget announcement's impact on the Indian stock market}


\author*[1]{\fnm{Mridul} \sur{Patel}}\email{mridul.patel@student.rmit.edu.au}

\author[1]{\fnm{G.} \sur{Verma}}\email{geetika.verma@rmit.edu.au}
\equalcont{These authors contributed equally to this work.}

\author[1]{\fnm{A.} \sur{Eberhard}}\email{andy.eberhard@rmit.edu.au}
\equalcont{These authors contributed equally to this work.}

\author[1]{\fnm{A.} \sur{Rao}}\email{asha.rao@rmit.edu.au}
\equalcont{These authors contributed equally to this work.}

\author[2]{\fnm{P.} \sur{Kumar}}\email{maths.pk@gmail.com, pankajkumar@nith.ac.in}

\affil[1]{\orgdiv{School of Mathematical Sciences}, \orgname{RMIT University}, \orgaddress{\street{124, LaTrobe Street}, \city{Melbourne}, \postcode{3000}, \state{VIC}, \country{Australia}}}

\affil[2]{\orgdiv{Department of Mathematics and Scientific Computing}, \orgname{National Institutr of Technology Hamirpur}, \orgaddress{\street{}\city{Hamirpur}, \postcode{177005}, \state{Himachal Pradesh}, \country{India}}}

% \affil[2]{\orgdiv{Department}, \orgname{Organization}, \orgaddress{\street{Street}, \city{City}, \postcode{10587}, \state{State}, \country{Country}}}

% \affil[3]{\orgdiv{Department}, \orgname{Organization}, \orgaddress{\street{Street}, \city{City}, \postcode{610101}, \state{State}, \country{Country}}}


\abstract{The stock market closely monitors macroeconomic policy announcements, such as annual budget events, due to their substantial influence on various economic participants. These events tend to impact the stock markets initially before affecting the real sector. Our study aims to analyze the effects of the budget on the Indian stock market, specifically focusing on the announcement for the year 2024. We will compare this with the years 2023, 2022, and 2020, assessing its impact on the NIFTY50 index using average abnormal return (AAR) and cumulative average abnormal return (CAAR) over a period of ±15 days. This study utilizes an innovative approach involving the fractal interpolation function, paired with fractal dimensional analysis, to study the fluctuations arising from budget announcements. The fractal perspective on the data offers an effective framework for understanding complex variations. }




\maketitle

\section{Introduction}\label{Intro}
The announcement of a national budget is a pivotal event that often significantly impacts the stock market. Investors closely scrutinize these announcements as they provide insights into the government's fiscal policies, including taxation, public spending, and economic reforms. A budget that signals increased government spending in sectors such as infrastructure, healthcare, or manufacturing can lead to a positive reaction in the stock market, as it is perceived to boost economic growth and corporate profitability. Conversely, investors may view measures that increase taxes or reduce spending negatively, leading to a sell-off in the stock market.

The stock market's response to budget announcements also varies based on specific policy details and their potential impact on different sectors. For example, a budget that increases capital expenditure is likely to benefit infrastructure and real estate stocks, while changes in tax policy could impact investor sentiment across the board. Historical trends have shown that markets tend to react favourably to growth-oriented budgets that focus on economic stimulus and reforms, whereas budgets perceived as lacking in substantial economic incentives or imposing heavier tax burdens can lead to market downturns. Singhvi \cite{singhvi2014impact}
explained that budget announcements are crucial in shaping market dynamics and influencing investment decisions. 

This research analyzes the impact of budget announcements on the Indian stock market. We examine the NIFTY50, commonly known as NIFTY, which serves as India’s primary stock market index on the National Stock Exchange (NSE). To assess the impacts of the budget announcement, we employ the Capital Asset Pricing Model (CAPM) to calculate the expected and actual returns of the stock market both before and after the budget announcement.

To quantitatively assess these varied market responses to budget announcements, financial metrics such as the Average Abnormal Return (AAR) and Cumulative Average Abnormal Return (CAAR) are often employed, offering deeper insights into the immediate and lasting impacts of fiscal policies on stock performance.

The Average Abnormal Return (AAR) and Cumulative Average Abnormal Return (CAAR) of NIFTY50 stocks during budget announcement periods have been utilized to evaluate market reactions to fiscal policies. The AAR quantifies the average deviation of a stock’s actual returns from its expected returns, capturing the immediate effects of the budget on stock prices. In contrast, the CAAR aggregates these abnormal returns over a specified period, offering a broader perspective on market sentiment and the lasting impact of the budget announcement. The analysis of AAR and CAAR for NIFTY50 stocks is considered crucial for understanding whether the market’s response to the budget was favorable or adverse. Such insights are deemed valuable for investors and policymakers, enabling more accurate predictions of market trends, informed investment decisions, and the formulation of future fiscal strategies.

% Calculating the Average Abnormal Return (AAR) and Cumulative Average Abnormal Return (CAAR) of NIFTY50 stocks surrounding budget announcements offers valuable insights into how the market reacts to fiscal policies. The AAR captures the average discrepancy between a stock's actual returns and its expected returns, reflecting the immediate effects of the budget on stock prices. In contrast, the CAAR accumulates these abnormal returns over a specified period, providing a comprehensive overview of market sentiment and the lasting impact of the budget announcement. Analyzing AAR and CAAR for the NIFTY50 is essential for understanding whether the market's reception of the budget was positive or negative. This analysis is vital for both investors and policymakers, as it aids in predicting market trends, making informed investment decisions, and developing future fiscal strategies.

Numerous authors have explored the impact of union budget announcements from various angles. Anil \cite{soni2010reaction} examined the stock market's response to monetary policy and budget announcements over a decade-long period. Additionally, the impact of budget announcements on both broad and sectoral indices of the Indian stock market is discussed in \cite{deepak2014event}. The work in \cite{patelimpact} analyzed the effects of budget announcements on the S\&P CNX NIFTY in terms of returns and volatility, employing paired t-tests and F-tests on returns during pre- and post-budget announcement trading days. Jain and Mahapatra \cite{jain2024testing} tested market efficiency by analyzing sectoral reactions to the union budget announcement. Goyal conducted an event study on the defense sector's response to the interim union budget announcement for 2024 in \cite{goyal2024analyzing}.

% This study focuses on examining the intricate trend of AAR and CAAR from a fractal perspective within a specified time frame. Fractals are intricate geometric forms that exhibit self-similarity, meaning each component replicates the overall pattern on a smaller scale. Unlike conventional geometric figures, fractals manifest in nature, seen in mountain shapes, coastlines, tree branches, and even the patterns of blood vessels \citep{mandelbrot1982fractal}. Benoît B. Mandelbrot popularized the concept of fractals in his influential work in \citep{mandelbrot1982fractal}, which demonstrated how mathematically describing and analyzing many irregular structures became achievable through fractal geometry. This novel viewpoint facilitated more precise modelling of natural phenomena, surpassing the limitations of traditional Euclidean geometry \citep{feder1988fractals}.

This study aims to investigate the intricate trends of Average Abnormal Returns (AAR) and Cumulative Average Abnormal Returns (CAAR) from a fractal perspective within a defined time frame.  Fractal interpolation function techniques offer effective deterministic representations of complex phenomena. Pioneers in this field, such as Barnsley \cite{barnsley1986fractal} and Hutchinson \cite{hutchinson1981fractals}, utilized fractal functions to interpolate datasets. Fractal interpolants can be applied to any continuous function defined on a real compact interval. This approach represents a significant advancement in approximation techniques, as all classical methods of real-data interpolation can be generalized through fractal methods \cite{chand2006generalized, navascues2004generalization}. The fractal interpolation function is used to analyze the fluctuations and complexity in the data set. Recently, Verma and Kumar \cite{verma2023fractal} analysed the stock prices of some companies after mergers and acquisitions. They also used the fractal interpolation function to analyze the S\&P BSE Sensex in India in \cite{kumar2024alpha}. The fractal dimension study of the COVID-19 data set through the fractal interpolation function presented by Agrawal and Verma in \cite{agrawal2023dimensional}.

Fractal interpolation provides a method to calculate the complexity of a trend through various fractal dimensions, such as box dimension \cite{verma2019revisit}, Hausdorff dimension \cite{fernandez2014fractal}, and Assoud dimension \cite{fraser2018assouad}. The fractal dimension serves as a tool to measure the degree of irregularity or roughness in time series data, offering valuable insights into the underlying complexity of return patterns. A higher fractal dimension indicates greater complexity and irregularity, suggesting that asset returns (and consequently the Average Abnormal Returns (AAR) and Cumulative Average Abnormal Returns (CAAR)) exhibit more unpredictable and chaotic behaviour. Conversely, a lower fractal dimension implies that return patterns are smoother and less complex, making them potentially simpler to model using traditional approaches.

By applying the fractal dimension to AAR and CAAR, we can gain a deeper understanding of how returns respond to market events while considering the inherent complexity and nonlinear dynamics of financial markets. This approach can enhance the accuracy of event study results and provide more robust insights into the impact of budget announcements on the Indian stock market.

 % Some of the researchers have analyzed the impact of budget announcement on different stock markets**cite some papers** using regression analysis.

 The rest of the paper is structured as follows: In Section \ref{ff}, we will discuss the construction of the fractal interpolation function and define the fractal dimension. A case study analyzing the impact of the budget announcement of the year 2024 on the Indian stock market is presented in Section \ref{cs}. The fractal analysis and box-dimensional analysis are presented in subsections \ref{FA} and \ref{BDA}, respectively. Finally, subsection \ref{RA} explains the result analysis of the impact of the budget announcement.

% Fractals have diverse applications across a range of fields including physics, biology, finance, computer graphics, geology, and geography see \citep{barnsley2013fractals,goldberger2002fractal,mandelbrot1997fractals,ebert2002texturing,mandelbrot1982fractal}. The study of fractals provides a mathematical framework that helps us understand the complexity of the irregular shapes and graphs.

In the following section, we will define the fractal interpolation function with the constant scaling factor and box dimension of the fractal function. 
\section{Fractal interpolation function}\label{ff}

% This section provides a few useful definitions and results related to contraction mappings and iterated function systems (IFS). 
 
% \subsection{Iterated function systems (IFS)}
Consider the data $\lbrace (x_i,y_i)  \in A\times \mathbb{R}:i=0,1,\ldots,P\rbrace$, where $A=[x_0,x_P]$ and $A_p=[x_{p-1},x_p]$, and the abscissae are strictly increasing. We have $P$ contractive homeomorphisms $l_p : A \rightarrow A_p$, each defined by $l_p(x) = a_p x+b_p$. The parameters $a_p$ and $b_p$ are determined in \cite{barnsley1986fractal}, based on the following conditions.
\begin{align}\label{affine}
    l_p(x_0)=x_{p-1},~~~l_p(x_P)=x_p, ~~\forall~ p=1,2,\ldots,P.
\end{align}
Define $\mathcal{F}_p: K (:= A \times \mathbb{R}) \rightarrow \mathbb{R}$ as the functions that are continuous in the first argument and contractive in the second argument, subject to the conditions $\mathcal{F}_p(x_0, y_0)= y_{p-1}$ and $\mathcal{F}_p(x_P, y_P )=y_p$, fulfilling for $0<t_p<0$ and $~(x,y), (x,y^{\prime})\in A\times\mathbb{R}$
\begin{align}\label{F_p}
  |\mathcal{F}_p(x,y)-\mathcal{F}_p(x,y^{\prime})|\leq t_p |y-y^{\prime}|.
\end{align}
Define the map $\Phi_p : K (:= A \times \mathbb{R})\rightarrow K (:= A \times \mathbb{R})$ by $\Phi_p(x, y) = (l_p(x), \mathcal{F}_p(x,y))$ for all $p\in \mathbb{N}$. Then, the system defined by 
\begin{align}\label{IFS}
\lbrace K;\Phi_p: p=1,2,\ldots, P \rbrace
\end{align}
is called the iterated function system with a finite number of contraction mappings, and the attractor of the IFS is a unique invariant set $A$ such that $A=  \displaystyle \bigcup^{P}_{n=1} \Phi_p (A)$. The set $A$ is the graph of a continuous function $g: A \rightarrow \mathbb{R} $ which satisfies $g(x_i)=y_i, i=0, 1,2,\ldots, P$ , which satisfies the functional equation, 
% Let $H(K)$ be the family of all non-empty compact subsets of $K$, consider a self map $W$ on $H(K)$ defined by $W(A)=  \displaystyle \bigcup^{N}_{n=1} \Phi_n (A)$. From the uniqueness and existence of fractal interpolation function described in \cite{barnsley1986fractal}, the self-map $W$ has a unique compact set $G$ such that $W(G)=G$. This set $G$ is called the attractor of the iterated function system \eqref{IFS}, and $G$ is the graph of a continuous function $f: I \rightarrow \mathbb{R} $ which satisfies $f(x_n)=y_n, n=1,2,\ldots, N.$ The function $f$ is called the Fractal Interpolation Function (FIF) or simply fractal function corresponding to the IFS \eqref{IFS}, which satisfies the functional equation, 
\begin{align}\label{f_Fn}
g(l_p(x))=\mathcal{F}_p(x,g(x)),~~~~\forall ~x\in A_p,~p\in P.
\end{align}
The most common IFS is constructed using the following maps $l_p(x)$ and $\mathcal{F}_p(x,y)$, which are defined as follows:
% The FIF is constructed from the following equation associated with the IFS \eqref{IFS}:
\begin{align}\label{ln_Fn}
l_p(x) = a_p x+b_p~~~~\mathcal{F}_p(x,y)= \alpha_p y + q_p(x),~(x,y)\in A\times \mathbb{R},~~p\in \mathbb{N},
\end{align}
where  $a_p \in \mathbb{R}$ and $q_p:A\rightarrow \mathbb{R}$ is a continuous functions satisfying $$q_p(x)=g(l_p(x))-\alpha_p b(x).$$ The function $q_p$ meets the endpoint conditions with $F_p$, where $b:\rightarrow \mathbb{R}$ is a base function satisfying the conditions $b(x_0)=y_0,b(x_P)=y_P$. Moreover, $\alpha_i's$ are the vertical scaling factors of the mappings $\Phi_p's$, with $\alpha_p$ falling within the range of $(-1, 1)$. The scale vector $\alpha=(\alpha_1,\alpha_2,\ldots,\alpha_P)$ is known as the scale vector corresponding to the IFS \eqref{IFS}. The function $g: A\rightarrow \mathbb{R}$ is the interpolation function.

The IFS $\lbrace K;\Phi_p: p=1,2,\ldots, P \rbrace$ has a unique attractor given by the graph of the continuous function $g$ by \eqref{fp}, denoted by $g^{\alpha}$ and satisfies the self-referential equation
\begin{align*}
    g^{\alpha}(x)=g(x)+\alpha_p (g^{\alpha}-b)(l^{-1}_{p}(x)), ~\mbox{for}~x\in A_p,~\mbox{and}~n\in \lbrace p\in 1,2,\ldots,P \rbrace.
\end{align*}

The function $g^{\alpha}$ is called $\alpha$-FIF. Readers may refer to the following literature for in-depth information about $\alpha$-FIF: \citep{wang2013fractal, barnsley1989hidden, yun2019hidden, vijender2019approximation, navascues2005fractal, navascues2007non, akhtar2017box, agathiyan2022construction}. The FIF with scaling factors is more versatile and offers greater flexibility and degrees of freedom in approximation. The application of the Banach contraction principle ensures the existence and uniqueness of the FIF, as guaranteed by the following theorem.
% \begin{align}\label{FIF}
% f(x)=\alpha_n (l_n^{-1}(x))f(l_n^{-1}(x))+q_n(l_n^{-1}(x)),~~x\in I_n,~n\in N, 
% \end{align}
% is called the $\alpha$-fractal interpolation function with constant scaling factors.

\begin{theorem}[Theorem 1, \citep{barnsley1986fractal}]\label{fp}
Consider the IFS $\lbrace K;\Phi_p: p=1,2,\ldots, P \rbrace$ defined in \eqref{IFS}. Let $\mathcal{C}$ represent the space of continuous functions $h^\prime:A\rightarrow \mathbb{R}$ with the conditions $h(x_0)=y_0$ and $h(x_P)=y_P$, equipped with the uniform metric $d(g,h)=\max\lbrace|h(x)-h^\prime(x)|:x\in A \rbrace$. The Read-Bajaraktarevic operator $\mathcal{T}$ is defined on the complete metric space $(\mathcal{C},d)$ as $(\mathcal{T}h)(x)=\mathcal{F}_p(l_p^{-1}(x),h(l_p^{-1}(x))),~\forall~x \in A_p,~p=1,2,\ldots,P$. Then
\begin{enumerate}
    \item The IFS possesses a unique attractor $G_g$, which represents the graph of a continuous function $g:A\rightarrow \mathbb{R}$ that interpolates the data $\lbrace (x_i,y_i)  \in A\times \mathbb{R}:i=0,1,\ldots,P\rbrace$.
    \item The operator $\mathcal{T}$ is a Banach contraction on $(\mathcal{C},d)$ and $g$ is its fixed point and satisfies the fixed point equation,
\end{enumerate}
\begin{align}\label{FIF}
g(x)=\alpha_p g(l_p^{-1}(x))+q_p(l_p^{-1}(x)),~~x\in A_p,~p\in P. 
\end{align}
The function $g$, which is continuous and satisfies equation (\ref{FIF}), is referred to as the FIF associated with the set of points $\lbrace (x_i,y_i)  \in A\times \mathbb{R}:i=0,1,\ldots, P\rbrace$.
\end{theorem}

\subsection{Fractal dimension}
The fractal dimension is important for understanding the complexity of fractals. The box-counting dimension is used to characterize the scaling properties of fractals by demonstrating how the detail or complexity of the fractal changes as it is zoomed in on. The box-counting dimension is defined as follows:
\begin{definition}[Box Dimension \citep{navascues2005fractal}] 
Let $Z$ be a nonempty bounded subset of the metric space $(X,d)$. The box dimension of $Z$ is defined as 
\begin{align*}
\dim_B Z=\lim_{\epsilon\to 0} \frac{\log N_{\epsilon}(Z)}{-\log \epsilon},
\end{align*}
where $\log N_{\delta}(Z)$ represents the minimum number of sets with a diameter at most $\epsilon$ that can cover $Z$, assuming the limit exists. If the limit doesn't exist, then the upper and lower box dimensions are respectively defined as follows:
\begin{align*}
    \overline{\dim}_B Z&= \limsup_{\epsilon\to 0} \frac{\log N_{\epsilon}(Z)}{-\log \epsilon},\\
        \underline{\dim}_B Z&= \liminf_{\epsilon\to 0} \frac{\log N_{\epsilon}(Z)}{-\log \epsilon}.
\end{align*}
\end{definition}
% \begin{theorem}\label{dim}\citep{barnsley1989hidden} Let $\Omega=(x_0,x_2,\ldots,<x_P)$ be a partition of $A=[x_0,x_P]$ satisfying $x_0<x_1<,\ldots,x_P$ and let $\alpha=(\alpha_1,\alpha_2,\ldots,\alpha_P)\in (-1,1)^P$. Assume that $g$ and $b$ are Lipschitz functions defined on $A$ with $b(x_0)=g(x_0)$ and $b(x_P)=g(x_P)$. If the data points $\lbrace (x_i,g(x_i)):i=0,1,\ldots,P \rbrace$ are not collinear, then 
% \begin{align*}
% \dim_B (G_r(g^{\alpha}_{\Omega,b}))=
% \begin{cases}
%      &D,~ if \sum^{P}_{i=1} |\alpha_i|>1; \\
%      &1,~ otherwise,
% \end{cases}
% \end{align*}
% where $(G_r(g^{\alpha}_{\Omega,b}))$ denotes the graph of $g^{\alpha}_{\Omega,b}$  and $D$ is the unique positive solution of the equation given as 
% \begin{align*}
%     \sum^{P}_{i=1} |\alpha_i| a_i^{D-1}=1.
% \end{align*}
% \end{theorem}

In the following section, we will use the fractal interpolation function and the box dimension defined in section \ref{ff} to examine the fluctuations in the stock index NIFTY50.   

\section{Case study}\label{cs}

This case study explores the impact of the Indian government's annual budget announcements on the stock market, with a focus on fluctuations within a stock index. The Indian stock market is known for its high sensitivity to budget proposals, which can significantly influence economic policy, taxation, and government spending priorities. Our findings suggest that the market typically experiences increased volatility before the budget, with reactions, whether positive or negative, depending on how well the budget proposals align with market expectations.

In this research, we analyze the NIFTY50, commonly known as NIFTY, which serves as India's primary stock market index on the National Stock Exchange (NSE). The NIFTY represents the weighted average of the 50 largest and most liquid Indian companies across 13 sectors, making it a crucial indicator of the Indian equity market and the overall economy. The NIFTY, a well-diversified index comprising 50 stocks that span 24 sectors of the economy, assesses the impact of the union budget announcement on these stocks within a 31-day period. As of September 3, 2024, the NIFTY stocks represent approximately 66\% of the total free float market capitalization of all stocks traded on the NSE.

The primary objective of this study is to examine the effect of the union budget on the NIFTY index of the national stock exchange, focusing specifically on average abnormal returns (AAR) and cumulative average abnormal returns (CAAR). Furthermore, the study will investigate the impact of the union budget in the periods before and after budget announcements.

To assess the influence of the budget announcement on NIFTY50, we will employ the fractal interpolation method. This approach provides a distinctive perspective for data analysis, as fractals exhibit self-similarity and involve a scaling factor. This method acts as an advanced tool for understanding trends and forecasting future movements based on historical data. In addition, we will apply box dimension analysis to evaluate the complexity of the trend across various scaling factors using the fractal interpolation function.

This research utilized data from India's National Stock Exchange (NSE) stock index. This analysis examines a period of 31 trading days surrounding the 2024 union budget. This timeframe is divided into three segments: 15 trading days prior to the budget day, the budget day itself, and 15 trading days following the budget day. The dataset is available at the website \url{www.nseindia.com}.
 % \vspace{5cm}

\subsection{Methodology and algorithm to analyze the impact of budget announcement}

\begin{figure}[htp]
    \centering
    \includegraphics[width=\linewidth]{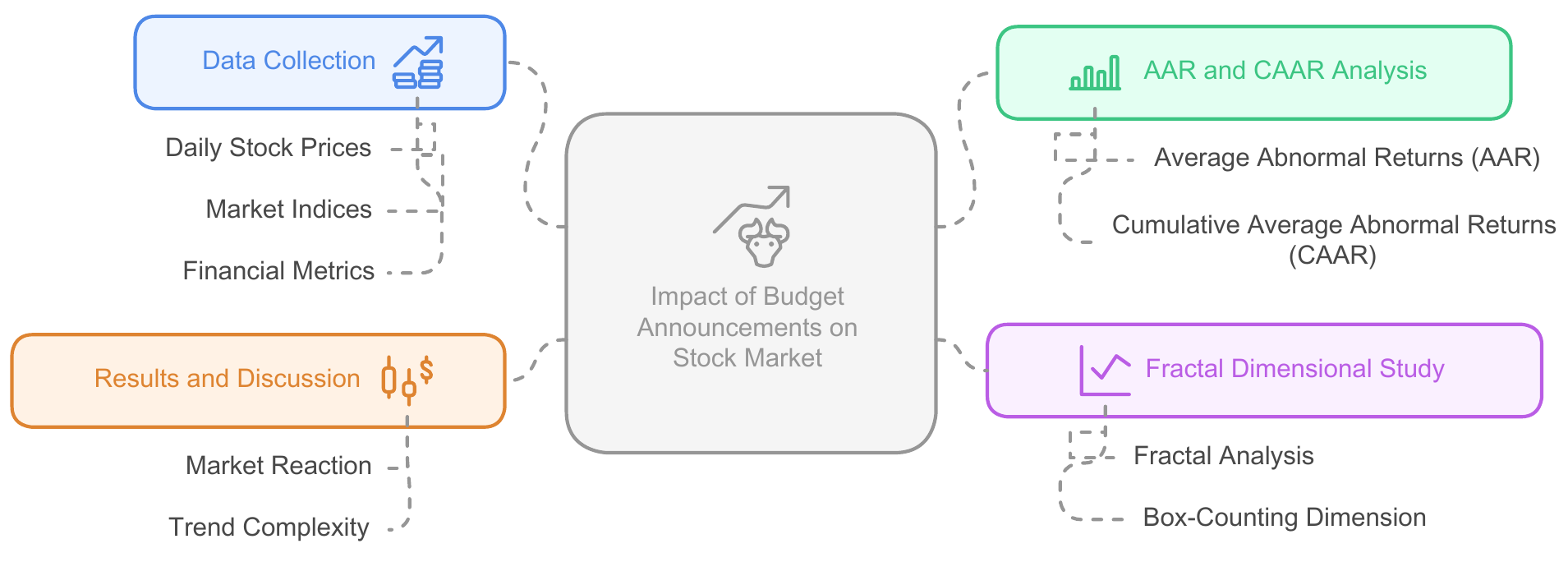}
    \caption{Methodology to analyze the impact of budget announcement}
    \label{fig:enter-label}
\end{figure}

\begin{figure}[!htbp]
    \centering
    \includegraphics[width=0.73\linewidth]{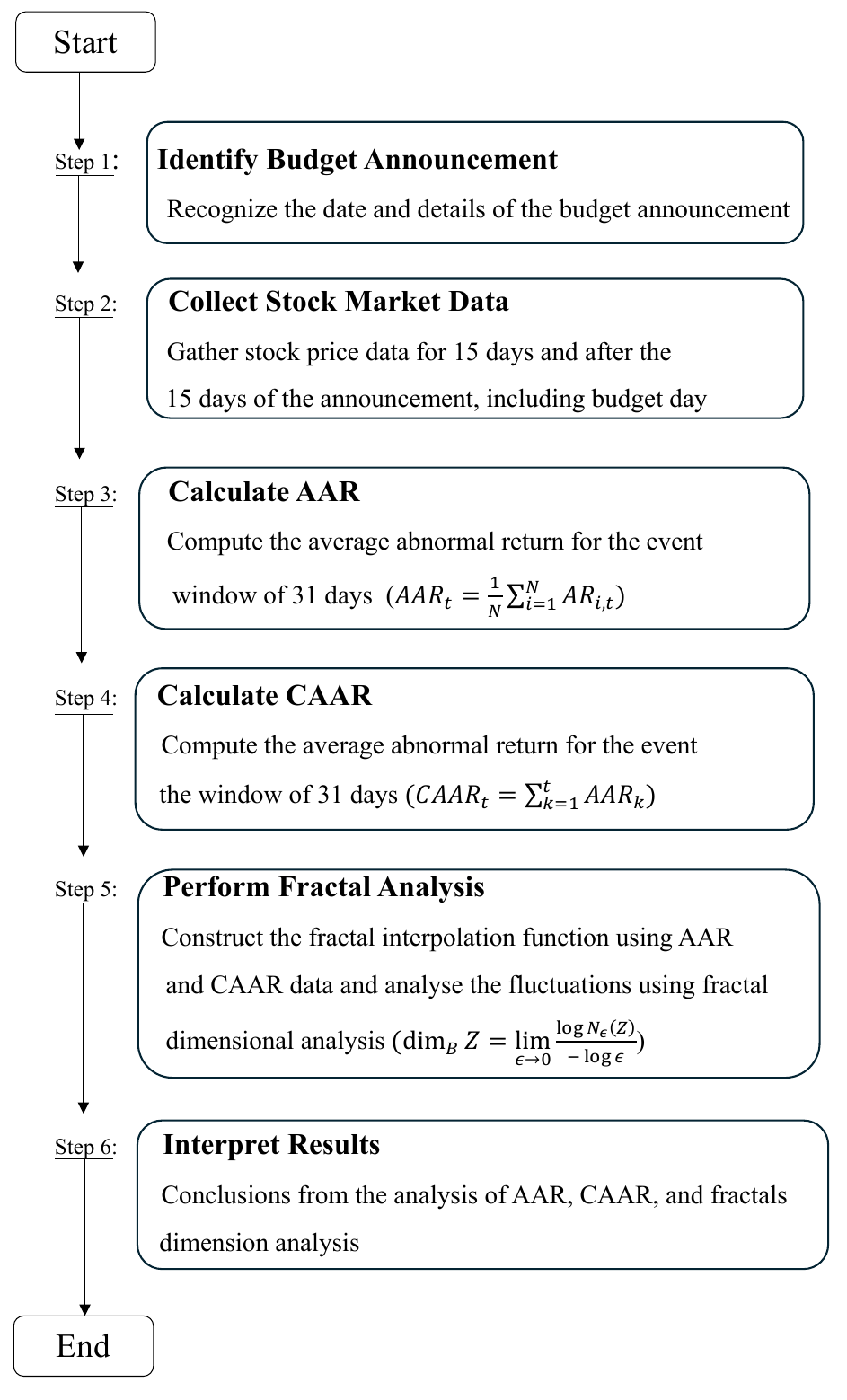}
    \caption{Algorithm to analyze the impact of the budget announcement on the stock market}
    \label{algo_budget}
\end{figure}

% \subsection{Objective of the study}

% The primary objective of this study is to examine the effect of the union budget on the NIFTY index of the national stock exchange, focusing specifically on average abnormal returns (AAR) and cumulative average abnormal returns (CAAR). Furthermore, the study will investigate the impact of the union budget in the periods before and after budget announcements.

% To assess the influence of the budget announcement on NIFTY50, we will employ the fractal interpolation method. This approach provides a distinctive perspective for data analysis, as fractals exhibit self-similarity and involve a scaling factor. This method acts as an advanced tool for understanding trends and forecasting future movements based on historical data. In addition, we will apply box dimension analysis to evaluate the complexity of the trend across various scaling factors using the fractal interpolation function.

% \subsection{Source of data}
% This research utilized data from India's National Stock Exchange (NSE) stock index. The dataset is available at the website www.nseindia.com.

% \subsection{Study period}
% This analysis examines a period of 31 trading days surrounding the 2024 union budget. This timeframe is divided into three segments: 15 trading days prior to the budget day, the budget day itself, and 15 trading days following the budget day.

\subsection{Statistical method}\label{stats}
The Capital Asset Pricing Model (CAPM) serves as a fundamental component of modern financial theory, offering a framework to understand the relationship between an asset’s expected return and its risk in relation to the broader market. According to CAPM, the expected return of an asset depends on the risk-free rate, the asset’s sensitivity to market fluctuations (measured by its beta), and the anticipated market return. The model posits that investors are rewarded for assuming systematic risk, asserting that the return on an asset is directly proportional to its exposure to the overall market.

To empirically estimate the model, linear regression is commonly utilized, wherein the asset's excess returns (returns above the risk-free rate) are regressed against the excess returns of the market portfolio. The slope coefficient of the regression line yields an estimate of the asset’s beta, which indicates its responsiveness to market movements, while the intercept reflects the asset's alpha, representing the portion of return that is unexplained by market fluctuations. By employing linear regression to estimate these parameters, CAPM provides a quantitative approach for assessing risk and predicting expected returns, thus enabling more informed decisions in asset pricing, portfolio management, and risk assessment. This empirical methodology is widely applied in finance to evaluate the validity of CAPM and to examine the performance of individual securities relative to the market.

\subsection*{Daily Returns}
The daily return of a stock indicates the change in its price from one day to the next. It serves as an important measure for assessing the fluctuations in the stock's value over time. The calculation of daily return can be expressed as follows:
\begin{equation}
\text{R}_t = \frac{\text{C}_t - \text{O}_{t}}{\text{O}_{t}}
\end{equation}
where:
\begin{align*}
R_t & = \text{Return on day t} \\
C_t & = \text{Closing price on day t} \\
O_t & = \text{Opening price on day t}
\end{align*}
\subsection*{Expected Return Using CAPM}
The Capital Asset Pricing Model (CAPM) is a widely recognized tool that is used to estimate the expected return on an asset, such as a stock or a stock index, by examining its risk in relation to the overall market. This model hinges on the interaction between the risk-free rate, the anticipated return of the market, and the asset's beta, which measures its sensitivity to market fluctuations, which is defined as follows:
\begin{equation}
R_i = R_f + \beta_i (R_m - R_f)
\end{equation}
\text{where:}
\begin{align*}
R_i & = \text{Expected return of the index} \\
R_f & = \text{Risk-free rate} \\
\beta_i & = \text{Beta of the index} \\
R_m & = \text{Return of the market index}
\end{align*}

\subsection*{Abnormal Return (AR)}
Abnormal Return refers to the discrepancy between the actual return of an asset, such as a stock or stock index, and its expected return, which is calculated using the Capital Asset Pricing Model (CAPM). Investors often use abnormal returns to assess how an asset has performed in comparison to what would normally be expected, taking into account its level of risk and the overall performance of the market.
\begin{equation}
AR_i = R_{\text{actual},i} - R_{\text{expected},i}
\end{equation}
\text{where:}
\begin{align*}
R_{\text{actual},i}&= \text{Actual return of stock}~ i,\\
R_{\text{expected},i}&= \text{Expected return of stock}~ i.
\end{align*}

\subsection*{Average Abnormal Return (AAR)}
Average Abnormal Return (AAR) is an important concept in financial analysis, especially in the context of event studies. It represents the average difference between the actual returns and the expected returns of a group of securities over a specific timeframe surrounding a particular event. 
\begin{equation}
AAR_t = \frac{1}{N} \sum_{i=1}^N AR_{i,t}
\end{equation}
\text{where:}
\begin{align*}
AR_{i,t}&= \text{The abnormal return for security i at time t}\\
N & = \text{Number of securities} \\
t & = \text{Number of days}
\end{align*}

\subsection*{Cumulative Average Abnormal Return (CAAR)}

The cumulative average abnormal return (CAAR) is a key metric employed in event studies to assess the overall abnormal returns of a group of securities over a designated time frame surrounding a specific event.
\begin{equation}
{CAAR}_t = \sum_{k=1}^t AAR_k.
\end{equation}

\subsection*{AAR and CAAR for NIFTY50}

Analyzing Average Abnormal Returns (AAR) and Cumulative Average Abnormal Returns (CAAR) is essential in finance, especially for assessing how specific events influence stock market performance. For indices like the Nifty 50, these metrics help investors and analysts understand how stock prices diverge from expected returns during a given event window. This analysis provides valuable insights into market reactions to news, earnings releases, or regulatory changes. By identifying abnormal returns, stakeholders can better gauge investor sentiment and market pricing efficiency, ultimately supporting more informed strategic decision-making.

This study utilizes 31 event windows, each spanning 31 days, encompassing 15 days before the event, the event day, and 15 days after the event. The research focuses on the NIFTY50 stock index on the National Stock Exchange of India, which lists the top 50 companies. By calculating the Average Abnormal Return (AAR) and Cumulative Average Abnormal Return (CAAR) during the event windows, the research aims to provide insights into the stock market's response to the union budget announcement. The AAR and CAAR values are computed in Table \ref{data} for the specified event window, employing the method outlined in Subsection \ref{stats}.

\begin{table}[!htbp]
    \centering
    \begin{adjustbox}{width=0.8\textwidth} 
    \begin{tabularx}{0.67\textwidth}{c*{5}{X}}\toprule
\textbf{S. No.} & \textbf{Day} & \textbf{$x_i$} & \textbf{AAR ($y_{1,i}$)} & \textbf{CAAR ($y_{2,i}$)} \\
\midrule
0 & -15 & 0.0000 &  0.00559 & 0.00559 \\
1 & -14 & 0.0333 &  0.00078 & 0.00637 \\
2 & -13 & 0.0667 & -0.00349 & 0.00287 \\
3 & -12 & 0.1000 & -0.00155 & 0.00132 \\
4 & -11 & 0.1333 & -0.00518 & -0.00386 \\
5 & -10 & 0.1667 &  0.00346 & -0.00039 \\
6 & -9  & 0.2000 &  0.00108 & 0.00069 \\
7 & -8  & 0.2333 &  0.00276 & 0.00345 \\
8 & -7  & 0.2667 & -0.00206 & 0.00139 \\
9 & -6  & 0.3000 &  0.00290 & 0.00429 \\
10 &-5  & 0.3333 & -0.00169 & 0.00260 \\
11 & -4  & 0.3667 & -0.00144 & 0.00115 \\
12 & -3  & 0.4000 &  0.00428 & 0.00544 \\
13 & -2  & 0.4333 &  0.00410 & 0.00954 \\
14 & -1  & 0.4667 &  0.00000 & 0.00954 \\
15 & 0  & 0.5000 & -0.00295 & 0.00659 \\
16 & +1  & 0.5333 & -0.00480 & 0.00179 \\
17 & +2  & 0.5667 & -0.00647 & -0.00468 \\
18 & +3  & 0.6000 &  0.00120 & -0.00348 \\
19 & +4  & 0.6333 & -0.00197 & -0.00545 \\
20 & +5  & 0.6667 & -0.00216 & -0.00762 \\
21 & +6  & 0.7000 &  0.00034 & -0.00727 \\
22 & +7  & 0.7333 &  0.00236 & -0.00490 \\
23 & +8  & 0.7667 &  0.00295 & -0.00195 \\
24 & +9  & 0.8000 & -0.00144 & -0.00339 \\
25 & +10  & 0.8333 & -0.00084 & -0.00424 \\
26 & +11  & 0.8667 & -0.00278 & -0.00702 \\
27 & +12  & 0.9000 &  0.00203 & -0.00499 \\
28 & +13  & 0.9333 & -0.00068 & -0.00567 \\
29 & +14  & 0.9667 &  0.00394 & -0.00172 \\
30 & +15 & 1.0000 &  0.00172 & 0.00000 \\ \bottomrule
    \end{tabularx}
    \end{adjustbox}
\caption{Average Abnormal Returns (AAR), and Cumulative Average Abnormal Returns (CAAR) for the index NIFTY50 (N=15). 0 represents the event day, -1 indicates one day before the event, +1 denotes one day after the event, and so on.}
    \label{data}
\end{table}

\begin{table}[htp]
    \centering
    \begin{adjustbox}{width=0.8\textwidth} 
    \begin{tabularx}{0.7\textwidth}{c*{4}{X}}\toprule
\textbf{S. No.} & \textbf{$x_i$} & \textbf{AAR ($y_{1,i}$)} & \textbf{CAAR ($y_{2,i}$)} \\
\midrule
0 & 0.0 &  0.00559 & 0.00559 \\
1 & 0.1 & -0.00155 & 0.00132 \\
2  & 0.2 &  0.00108 & 0.00069 \\
3  & 0.3 &  0.00290 & 0.00429 \\
4  & 0.4 &  0.00428 & 0.00544 \\
5  & 0.5 & -0.00295 & 0.00659 \\
6  & 0.6 &  0.00120 & -0.00348 \\
7  & 0.7 &  0.00034 & -0.00727 \\
8  & 0.8 & -0.00144 & -0.00339 \\
9  & 0.9 &  0.00203 & -0.00499 \\
10  & 1.0 &  0.00172 & 0.00000 \\ \bottomrule
    \end{tabularx}
    \end{adjustbox}
\caption{Average Abnormal Returns (AAR), and Cumulative Average Abnormal Returns (CAAR) for NIFTY50 (N=15) }
    \label{data_1}
\end{table}

\subsection{Fractal Analysis}\label{FA}

We will use the fractal interpolation function as described in Section \eqref{ff} to examine the fractal patterns of AAR and CAAR, as computed in Table \eqref{data}. Additionally, we will explore the box dimension analysis to comprehend the complexity of the fractal function across various values of $\alpha$. In this case, the germ function is taken as and the base function $b_1$ and $b_2$ are taken as $b_1=f_1(x^2), b_2=f_2(x^2)$, and the data set $\lbrace (x_i,y_{j,i}):i=0,1,\ldots,10, j=1,2 \rbrace$ is not collinear. The functions $f_1(x), f_2(x)$ are satisfying the data set $\lbrace (x_i,y_{1,i}):i=0,1,\ldots,10 \rbrace$ and $\lbrace (x_i,y_{2,i}):i=0,1,\ldots,10 \rbrace$ respectively.

% \vspace{2cm}
% \[
$$
f_1(x) =
\begin{cases}
-0.0714x + 0.00559 & \text{for } 0.0 \leq x < 0.1 \\
0.0263x - 0.00418 & \text{for } 0.1 \leq x < 0.2 \\
0.0182x - 0.00256 & \text{for } 0.2 \leq x < 0.3 \\
0.0138x - 0.00124 & \text{for } 0.3 \leq x < 0.4 \\
-0.0718x + 0.033 & \text{for } 0.4 \leq x < 0.5 \\
0.041x - 0.0234 & \text{for } 0.5 \leq x < 0.6 \\
-0.0086x + 0.00636 & \text{for } 0.6 \leq x < 0.7 \\
-0.0178x + 0.0128 & \text{for } 0.7 \leq x < 0.8 \\
0.0347x - 0.0292 & \text{for } 0.8 \leq x < 0.9 \\
-0.0031x + 0.00482 & \text{for } 0.9 \leq x \leq 1.0
\end{cases}
$$
% \]
\vspace{3cm}
$$
f_2(x) =
\begin{cases}
-0.0427x + 0.00559 & \text{for } 0.0 \leq x < 0.1 \\
-0.0063x + 0.00195 & \text{for } 0.1 \leq x < 0.2 \\
0.0360x - 0.00651 & \text{for } 0.2 \leq x < 0.3 \\
0.0115x + 0.00084 & \text{for } 0.3 \leq x < 0.4 \\
0.0115x + 0.00084 & \text{for } 0.4 \leq x < 0.5 \\
-0.1007x + 0.05694 & \text{for } 0.5 \leq x < 0.6 \\
-0.0379x + 0.01926 & \text{for } 0.6 \leq x < 0.7 \\
0.0388x - 0.03443 & \text{for } 0.7 \leq x < 0.8 \\
-0.0160x + 0.00941 & \text{for } 0.8 \leq x < 0.9 \\
0.0499x - 0.0499 & \text{for } 0.9 \leq x \leq 1.0
\end{cases}
$$

\begin{figure}[htbp]
    \centering
    \subfloat[$\alpha$-FIF for $\alpha=0.3$ for the AAR data set $(x_i,y_{1,i})$]{
        \includegraphics[width=0.4\textwidth]{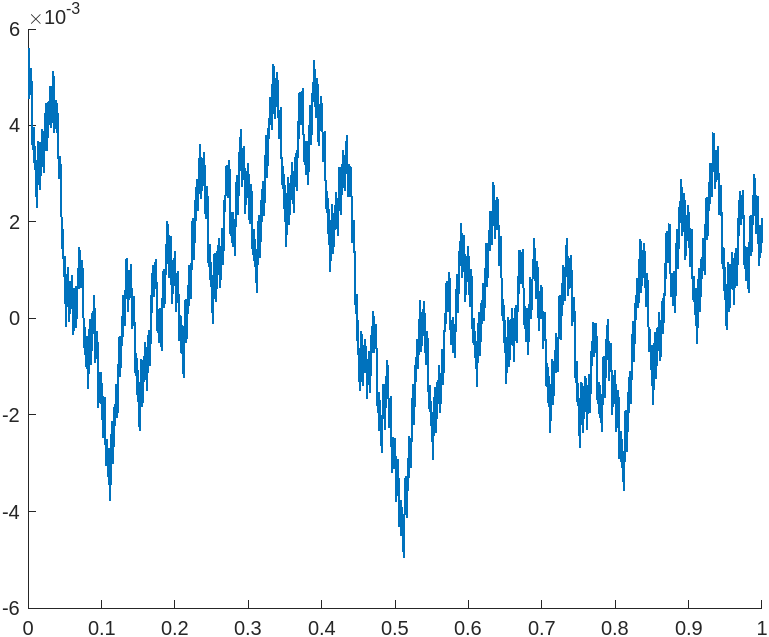}
        \label{AAR3}
    }
    \hspace{0.05\textwidth} % Adjust the horizontal space between figures
    \subfloat[$\alpha$-FIF for $\alpha=0.5$ for the AAR data set $(x_i,y_{1,i})$]{
        \includegraphics[width=0.4\textwidth]{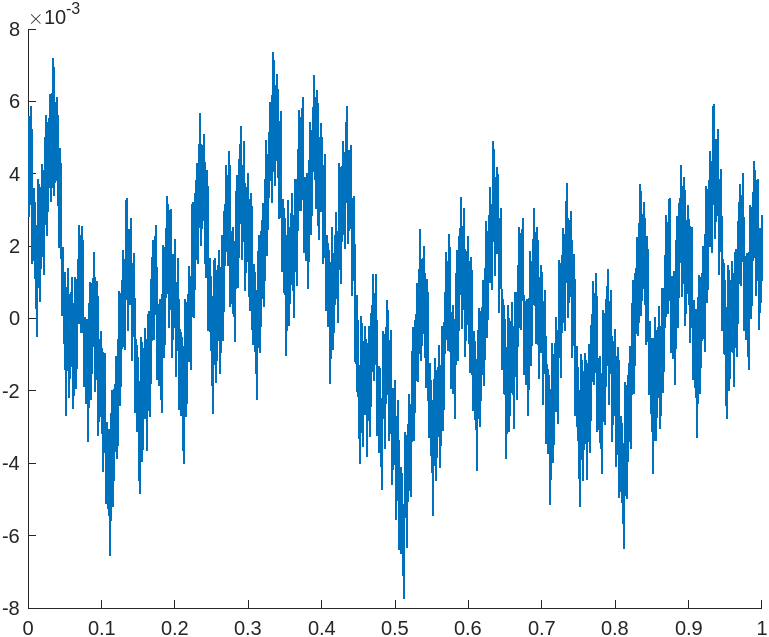}
        \label{AAR5}
    }
    \hspace{0.05\textwidth}
    \subfloat[$\alpha$-FIF  for $\alpha=(0.1, 0.4, 0.5, 0.6, 0.4, 0.3, 0.4,$ $ 0.5, 0.3, 0.1)$ for the AAR data set $(x_i,y_{1,i})$]{
        \includegraphics[width=0.4\textwidth]{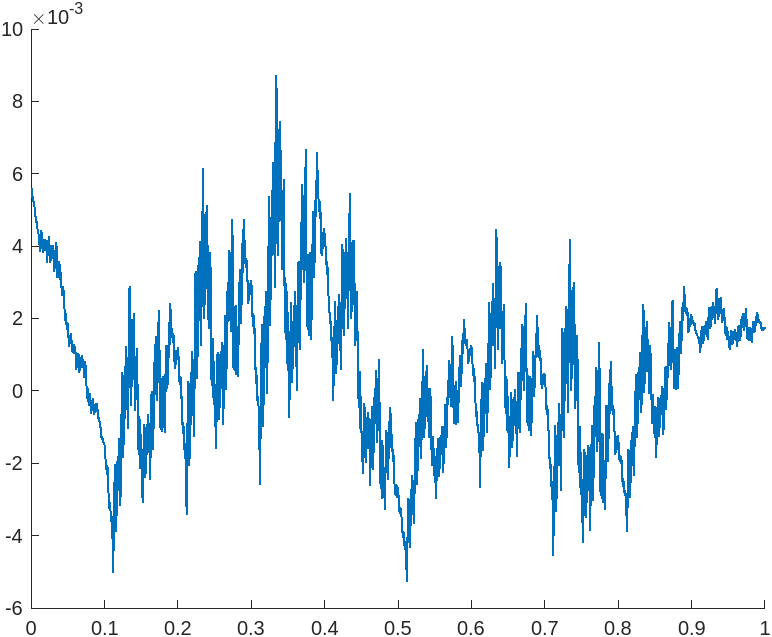}
        \label{AARMIX}
    }
        \hspace{0.05\textwidth}
    \subfloat[$\alpha$-FIF  for $\alpha=0 $, which represents classical interpolation of for the AAR data set $(x_i,y_{1,i})$]{
        \includegraphics[width=0.4\textwidth]{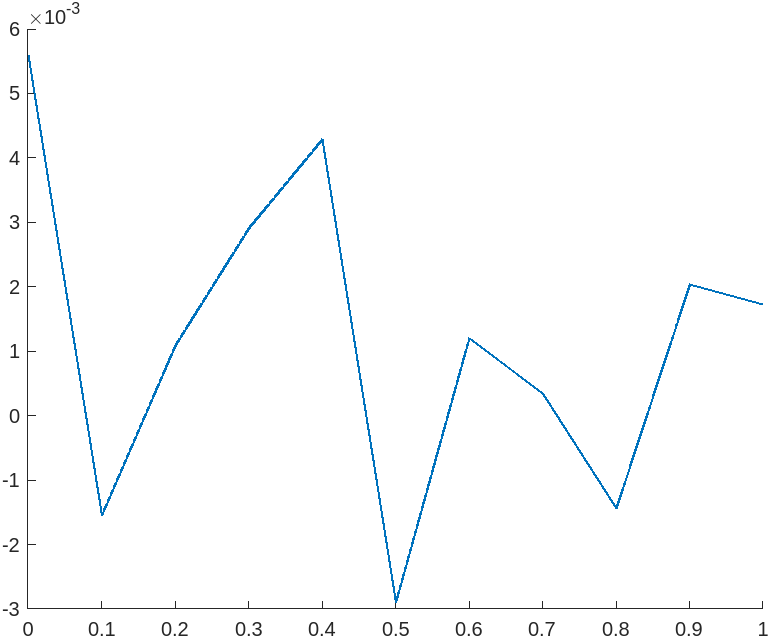}
        \label{AAR0}
    }
    \caption{$\alpha$-FIF for the AAR data set $(x_i,y_{1,i})$ for the different values of $\alpha$}
    \label{AAR}
\end{figure}

\begin{figure}[htbp]
    \centering
    \subfloat[$\alpha$-FIF for $\alpha=0.3$ for the CAAR data set $(x_i,y_{2,i})$]{
        \includegraphics[width=0.4\textwidth]{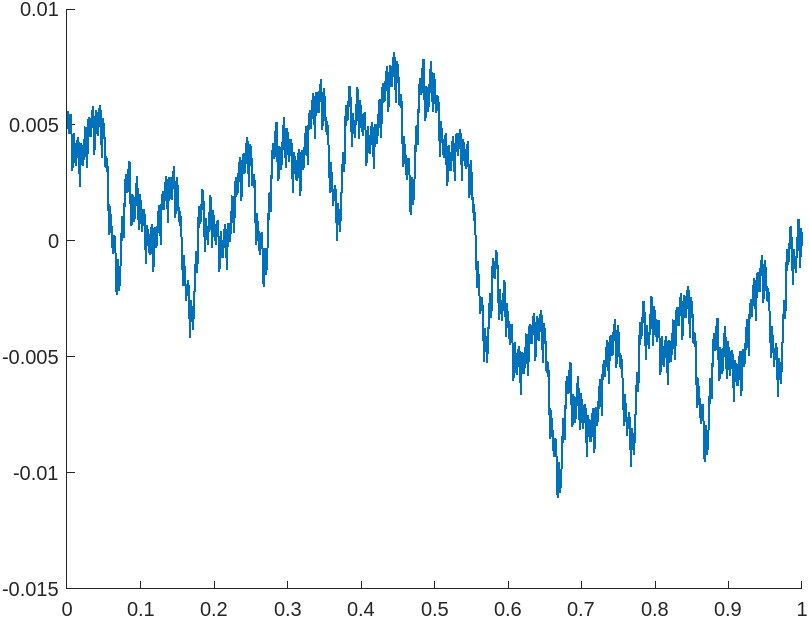}
        \label{CAAR3}
    }
    \hspace{0.05\textwidth} % Adjust the horizontal space between figures
    \subfloat[$\alpha$-FIF for $\alpha=0.5$ for the CAAR data set $(x_i,y_{2,i})$]{
        \includegraphics[width=0.4\textwidth]{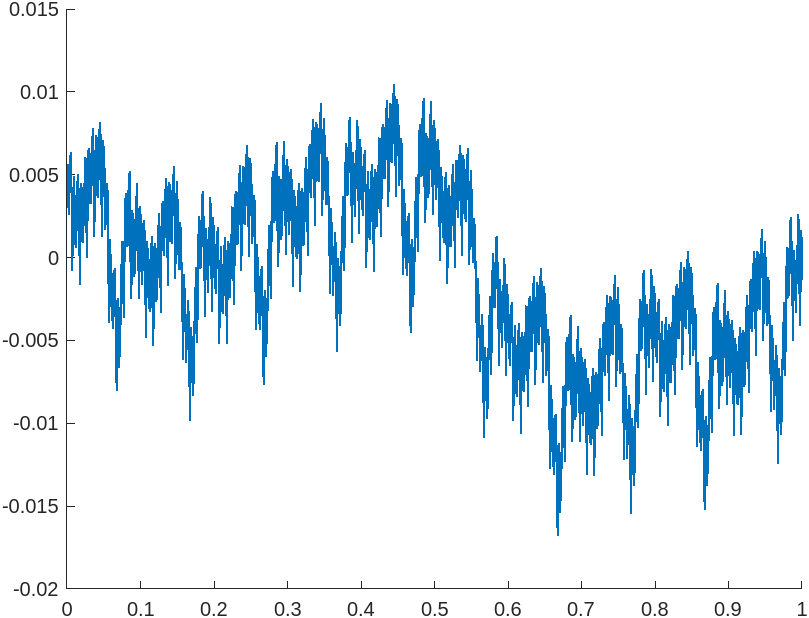}
        \label{CAAR5}
    }
    \hspace{0.05\textwidth}
    \subfloat[$\alpha$-FIF for $\alpha=(0.1, 0.4, 0.5, 0.6, 0.4, 0.3, 0.4, 0.5, 0.3, 0.1)$ for the CAAR data set $(x_i,y_{2,i})$]{
        \includegraphics[width=0.4\textwidth]{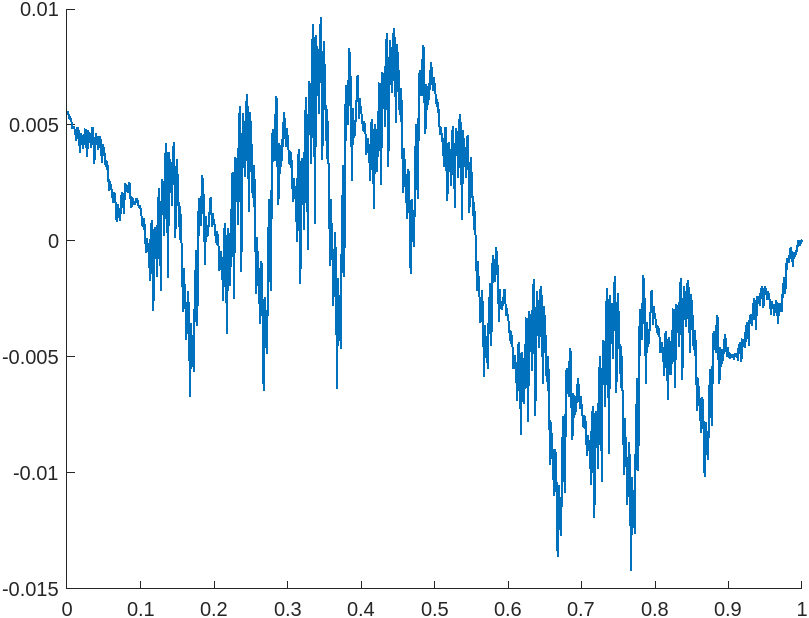}
        \label{CAARMIX}
    }
        \hspace{0.05\textwidth}
    \subfloat[$\alpha$-FIF for $\alpha=0$, which represents classical interpolation for the CAAR data set $(x_i,y_{2,i})$]{
        \includegraphics[width=0.4\textwidth]{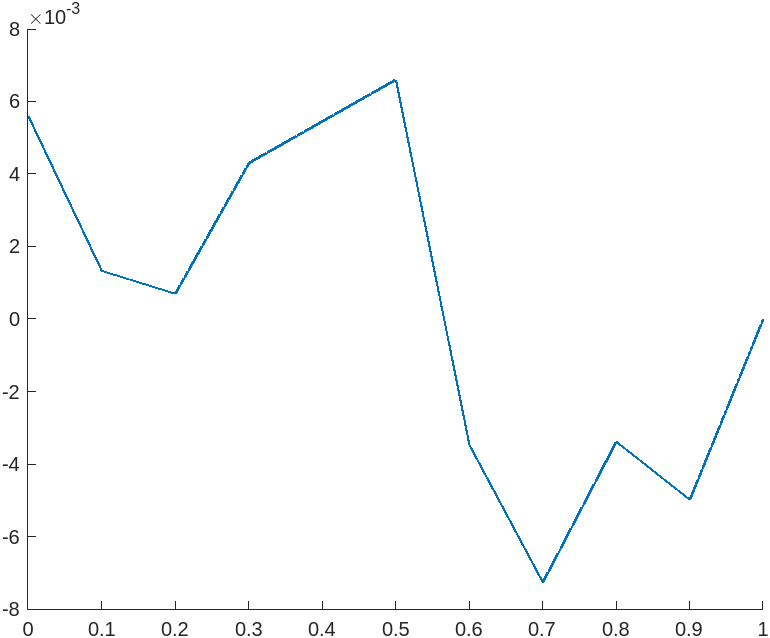}
        \label{CAAR0}
    }
    \caption{$\alpha$-FIF for the CAAR data set $(x_i,y_{2,i})$ for the different values of $\alpha$}
    \label{CAAR}
\end{figure}

  \vspace{0.2cm}
Fractal interpolation functions are constructed using the AAR and CAAR data sets for various values of the vertical scaling factor $\alpha$, as shown in figures \ref{AAR} and \ref{CAAR}. Variations in the vertical scaling factor $\alpha$ result in changes to the peaks and troughs of the FIFs graphs. An increase in the values of $\alpha$ leads to higher peaks and troughs, contributing to greater graph complexity. To quantify this complexity, the box dimension is calculated for different values of $\alpha$.

\subsection{Box Dimension Analysis}\label{BDA}
 Conducting a box dimension analysis allows us to investigate the complexity and variability inherent in the $\alpha$-FIF trends. We will compute the box dimension of AAR and CAAR presented in figures \ref{AAR} and \ref{CAAR} for various values of $\alpha$, as outlined below. Moreover, the fractal dimensional comparison of the impact of the budget announcements on the NIFTY50 is presented using the regression analysis presented in table \ref{BD}. An increase in the box dimension suggests a greater degree of fluctuations within the trend over the defined time interval.

\begin{table}[htp]
    \centering
    \resizebox{0.7\textwidth}{!}{ % Adjust the width of the table
        \begin{tabular}{c|c|c|c|c}
             \hline
            \multirow{1}{*}{Year} & \multicolumn{2}{c|}{AAR} & \multicolumn{2}{c}{CAAR} \\ 
                                   & $\alpha=0.3$ & $\alpha=0.5$ & $\alpha=0.3$ & $\alpha=0.5$ \\ \hline
            2024   & 1.468      & 1.595  & 1.378  & 1.498 \\ \hline
            2023  &  1.406      & 1.526   & 1.464 & 1.568 \\ \hline
            2022  &  1.465      & 1.546   & 1.377 & 1.506 \\ \hline
            2020  & 1.374 & 1.582   & 1.442  & 1.572 \\ \hline
        \end{tabular}
    }
    \caption{Box-counting dimensions of AAR and CAAR in the year 2024, 2023, 2022 and 2020 for $\alpha=0.3$ and $\alpha=0.5$}
    \label{BD}
\end{table}
% \vspace{3cm}
\begin{figure}[htp]
    \centering
    \includegraphics[width=0.65\linewidth]{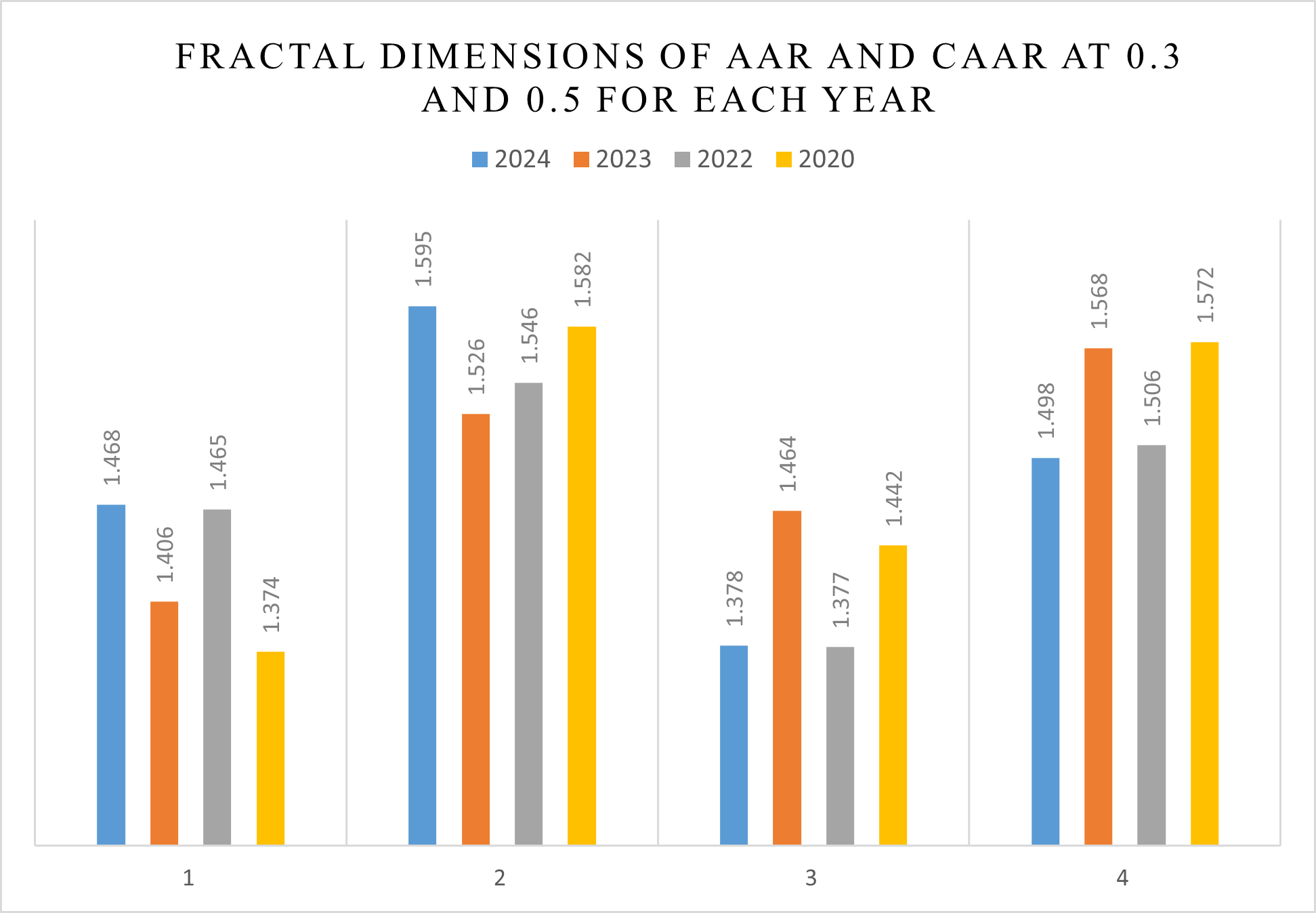}
    \caption{Fractal dimension comparison of AAR and CAAR at $\alpha=0.3$ and $\alpha=0.5$ for 2024, 2023, 2022 and 2020 }
    \label{FD_chart}
\end{figure}
The box-dimensions for the years 2024 and 2023 are calculated using the same scaling factor, $\alpha$. As illustrated in Table \ref{BD}, the Average Abnormal Returns (AAR) for 2024 surpass those for 2023 at both $\alpha=0.3$ and $\alpha=0.5$. This observation indicates that the AAR experienced more significant fluctuations during the event window of 2024 compared to that of 2023. Conversely, the Cumulative Average Abnormal Returns (CAAR) showed greater volatility during the event window of 2023 than in 2024. This implies that the long-term effects of the budget announcement on NIFTY50 were considerably more variable in 2023.

A similar analysis was conducted for the years 2022 and 2020. The year 2020 faced significant challenges due to the COVID-19 pandemic, which left it particularly vulnerable. India was among the countries heavily impacted by the effects of COVID-19. Therefore, while the analysis for 2020 does not delve into the pandemic itself or the accompanying lockdown restrictions, it's important to note that these factors significantly influenced the share market. We can compare the effects of union budget announcements across different years, including the year of the global pandemic. The box dimension comparison is shown in Table \ref{BD}, while the visual representation of the dimensions of AAR and CAAR for $\alpha=0.3$ and $\alpha=0.5$ is illustrated in Fig \ref{FD_chart}. 

\subsection{Result analysis}\label{RA}
The negative CAAR values following the budget announcement, in conjunction with Box Dimension analysis, reveal that the market's response was both unfavourable and complex. The negative abnormal returns observed on the NIFTY50 index indicate a generally pessimistic outlook among investors regarding the budget. The elevated Box Dimension suggests that the market's reaction was marked by considerable irregularity and unpredictability, likely reflecting a mix of external uncertainties and sector-specific influences.

This intricate CAAR pattern challenges the notion of a straightforward, rational market response to budgetary policy changes. Instead, the findings imply that dynamic and interconnected factors shaped the market's behaviour, indicating a more chaotic and less predictable market environment. Future research could delve into the specific sectors most impacted by the budget and assess whether this complexity aligns with historical budgetary events. Additionally, investigating the long-term effects of the budget on the market beyond the initial reaction period could yield further insights into the sustainability of this negative market response.

\section{Conclusion and future work}\label{con}
This research aims to investigate the impact of the union budget announcement on the National Stock Exchange (NSE) in India for the year 2024. The primary objective is to determine whether the stock index on the NSE reacts positively or negatively to the announcement. Additionally, the study analyzes the average abnormal return and cumulative average abnormal return before and after the budget announcement by the central government. The findings indicate that there is a significant negative trend in the cumulative average abnormal returns following the event date (the budget announcement).

The primary limitation of this study is its omission of other exogenous factors that influence stock market indices. Consequently, future research could expand the analysis by incorporating additional exogenous effects beyond budget announcements. Furthermore, various sectors may respond differently to tax impositions or concessions and exemptions. Thus, a sector-wise comparison would offer a deeper understanding of these dynamics. Additionally, we employed a constant scaling factor $\alpha$ to create the FIF, which can be adapted to a variable scaling factor.

% \bmhead{Acknowledgemen

 % \section*{Declarations}

% Some journals require declarations to be submitted in a standardised format. Please check the Instructions for Authors of the journal to which you are submitting to see if you need to complete this section. If yes, your manuscript must contain the following sections under the heading `Declarations':

%  \begin{itemize}
%  \item Funding
%  \item Conflict of interest/Competing interests (check journal-specific guidelines for which heading to use)
% % \item Ethics approval and consent to participate
% % \item Consent for publication
%  \item Data availability 
%  \url{https://www.abs.gov.au/}
% % \item Materials availability
% % \item Code availability 
%  \item Author contribution
%  \end{itemize}
\section*{Statements and Declaration} 
\textbf{Conflict of interest:} The authors declare that they have no conflict of interest.\\
\noindent\textbf{Ethical approval:} This article does not contain any studies with human participants or animals performed by any of the authors.\\
\noindent\textbf{Funding:} This research has no funding by any organization or individual.\\
\noindent\textbf{Data availability:} \url{https://www.nseindia.com/}

\bibliography{StockMarket}% common bib file


\begin{thebibliography}{24}
\ifx \bisbn   \undefined \def \bisbn  #1{ISBN #1}\fi
\ifx \binits  \undefined \def \binits#1{#1}\fi
\ifx \bauthor  \undefined \def \bauthor#1{#1}\fi
\ifx \batitle  \undefined \def \batitle#1{#1}\fi
\ifx \bjtitle  \undefined \def \bjtitle#1{#1}\fi
\ifx \bvolume  \undefined \def \bvolume#1{\textbf{#1}}\fi
\ifx \byear  \undefined \def \byear#1{#1}\fi
\ifx \bissue  \undefined \def \bissue#1{#1}\fi
\ifx \bfpage  \undefined \def \bfpage#1{#1}\fi
\ifx \blpage  \undefined \def \blpage #1{#1}\fi
\ifx \burl  \undefined \def \burl#1{\textsf{#1}}\fi
\ifx \doiurl  \undefined \def \doiurl#1{\url{https://doi.org/#1}}\fi
\ifx \betal  \undefined \def \betal{\textit{et al.}}\fi
\ifx \binstitute  \undefined \def \binstitute#1{#1}\fi
\ifx \binstitutionaled  \undefined \def \binstitutionaled#1{#1}\fi
\ifx \bctitle  \undefined \def \bctitle#1{#1}\fi
\ifx \beditor  \undefined \def \beditor#1{#1}\fi
\ifx \bpublisher  \undefined \def \bpublisher#1{#1}\fi
\ifx \bbtitle  \undefined \def \bbtitle#1{#1}\fi
\ifx \bedition  \undefined \def \bedition#1{#1}\fi
\ifx \bseriesno  \undefined \def \bseriesno#1{#1}\fi
\ifx \blocation  \undefined \def \blocation#1{#1}\fi
\ifx \bsertitle  \undefined \def \bsertitle#1{#1}\fi
\ifx \bsnm \undefined \def \bsnm#1{#1}\fi
\ifx \bsuffix \undefined \def \bsuffix#1{#1}\fi
\ifx \bparticle \undefined \def \bparticle#1{#1}\fi
\ifx \barticle \undefined \def \barticle#1{#1}\fi
\bibcommenthead
\ifx \bconfdate \undefined \def \bconfdate #1{#1}\fi
\ifx \botherref \undefined \def \botherref #1{#1}\fi
\ifx \url \undefined \def \url#1{\textsf{#1}}\fi
\ifx \bchapter \undefined \def \bchapter#1{#1}\fi
\ifx \bbook \undefined \def \bbook#1{#1}\fi
\ifx \bcomment \undefined \def \bcomment#1{#1}\fi
\ifx \oauthor \undefined \def \oauthor#1{#1}\fi
\ifx \citeauthoryear \undefined \def \citeauthoryear#1{#1}\fi
\ifx \endbibitem  \undefined \def \endbibitem {}\fi
\ifx \bconflocation  \undefined \def \bconflocation#1{#1}\fi
\ifx \arxivurl  \undefined \def \arxivurl#1{\textsf{#1}}\fi
\csname PreBibitemsHook\endcsname

%%% 1
\bibitem[\protect\citeauthoryear{Singhvi}{2014}]{singhvi2014impact}
\begin{barticle}
\bauthor{\bsnm{Singhvi}, \binits{A.}}:
\batitle{Impact of union budget on nifty}.
\bjtitle{Pacific Business Review International}
\bvolume{6}(\bissue{12}),
\bfpage{23}--\blpage{28}
(\byear{2014})
\end{barticle}
\endbibitem

%%% 2
\bibitem[\protect\citeauthoryear{Soni}{2010}]{soni2010reaction}
\begin{barticle}
\bauthor{\bsnm{Soni}, \binits{A.}}:
\batitle{Reaction of the stock market to union budget and monetary policy announcements}.
\bjtitle{Asia Pacific Journal of Research in Business Management}
\bvolume{1}(\bissue{2}),
\bfpage{155}--\blpage{175}
(\byear{2010})
\end{barticle}
\endbibitem

%%% 3
\bibitem[\protect\citeauthoryear{Deepak and Bhavya}{2014}]{deepak2014event}
\begin{barticle}
\bauthor{\bsnm{Deepak}, \binits{R.}},
\bauthor{\bsnm{Bhavya}, \binits{N.}}:
\batitle{An event study analysis of union budget announcement on broad and sectoral indices of indian stock market}.
\bjtitle{International Journal of Innovative Research and Development}
\bvolume{3}(\bissue{12}),
\bfpage{1}--\blpage{21}
(\byear{2014})
\end{barticle}
\endbibitem

%%% 4
\bibitem[\protect\citeauthoryear{Patel and Patel}{}]{patelimpact}
\begin{botherref}
\oauthor{\bsnm{Patel}, \binits{M.M.}},
\oauthor{\bsnm{Patel}, \binits{B.I.}}:
Impact of announcement of union budgets on the s\&p cnx nifty in terms of returns and volatility
\end{botherref}
\endbibitem

%%% 5
\bibitem[\protect\citeauthoryear{Jain et~al.}{2024}]{jain2024testing}
\begin{barticle}
\bauthor{\bsnm{Jain}, \binits{N.}},
\bauthor{\bsnm{Mahapatra}, \binits{S.K.}}, \betal:
\batitle{Testing market efficiency: An empirical study on sectoral reaction to union budget announcement}.
\bjtitle{South India Journal of Social Sciences}
\bvolume{22}(\bissue{3}),
\bfpage{104}--\blpage{114}
(\byear{2024})
\end{barticle}
\endbibitem

%%% 6
\bibitem[\protect\citeauthoryear{Goyal}{2024}]{goyal2024analyzing}
\begin{barticle}
\bauthor{\bsnm{Goyal}, \binits{T.}}:
\batitle{Analyzing the responses of defense sector stocks to interim union budget announcement 2024: An event study}.
\bjtitle{Sachetas}
\bvolume{3}(\bissue{3}),
\bfpage{35}--\blpage{42}
(\byear{2024})
\end{barticle}
\endbibitem

%%% 7
\bibitem[\protect\citeauthoryear{Barnsley}{1986}]{barnsley1986fractal}
\begin{barticle}
\bauthor{\bsnm{Barnsley}, \binits{M.F.}}:
\batitle{Fractal functions and interpolation}.
\bjtitle{Constructive approximation}
\bvolume{2},
\bfpage{303}--\blpage{329}
(\byear{1986})
\end{barticle}
\endbibitem

%%% 8
\bibitem[\protect\citeauthoryear{Hutchinson}{1981}]{hutchinson1981fractals}
\begin{barticle}
\bauthor{\bsnm{Hutchinson}, \binits{J.E.}}:
\batitle{Fractals and self similarity}.
\bjtitle{Indiana University Mathematics Journal}
\bvolume{30}(\bissue{5}),
\bfpage{713}--\blpage{747}
(\byear{1981})
\end{barticle}
\endbibitem

%%% 9
\bibitem[\protect\citeauthoryear{Chand and Kapoor}{2006}]{chand2006generalized}
\begin{barticle}
\bauthor{\bsnm{Chand}, \binits{A.}},
\bauthor{\bsnm{Kapoor}, \binits{G.}}:
\batitle{Generalized cubic spline fractal interpolation functions}.
\bjtitle{SIAM Journal on Numerical Analysis}
\bvolume{44}(\bissue{2}),
\bfpage{655}--\blpage{676}
(\byear{2006})
\end{barticle}
\endbibitem

%%% 10
\bibitem[\protect\citeauthoryear{Navascu{\'e}s and Sebasti{\'a}n}{2004}]{navascues2004generalization}
\begin{barticle}
\bauthor{\bsnm{Navascu{\'e}s}, \binits{M.A.}},
\bauthor{\bsnm{Sebasti{\'a}n}, \binits{M.V.}}:
\batitle{Generalization of hermite functions by fractal interpolation}.
\bjtitle{Journal of Approximation Theory}
\bvolume{131}(\bissue{1}),
\bfpage{19}--\blpage{29}
(\byear{2004})
\end{barticle}
\endbibitem

%%% 11
\bibitem[\protect\citeauthoryear{Verma and Kumar}{2023}]{verma2023fractal}
\begin{barticle}
\bauthor{\bsnm{Verma}, \binits{S.K.}},
\bauthor{\bsnm{Kumar}, \binits{S.}}:
\batitle{Fractal dimension analysis of stock prices of selected resulting companies after mergers and acquisitions}.
\bjtitle{The European Physical Journal Special Topics}
\bvolume{232}(\bissue{7}),
\bfpage{1093}--\blpage{1103}
(\byear{2023})
\end{barticle}
\endbibitem

%%% 12
\bibitem[\protect\citeauthoryear{Kumar et~al.}{2024}]{kumar2024alpha}
\begin{barticle}
\bauthor{\bsnm{Kumar}, \binits{A.}},
\bauthor{\bsnm{Verma}, \binits{S.K.}},
\bauthor{\bsnm{Boulaaras}, \binits{S.M.}}:
\batitle{On $\alpha$-fractal functions and their applications to analyzing the s\&p bse sensex in india}.
\bjtitle{Chaos, Solitons \& Fractals}
\bvolume{186},
\bfpage{115194}
(\byear{2024})
\end{barticle}
\endbibitem

%%% 13
\bibitem[\protect\citeauthoryear{Agrawal and Verma}{2023}]{agrawal2023dimensional}
\begin{barticle}
\bauthor{\bsnm{Agrawal}, \binits{E.}},
\bauthor{\bsnm{Verma}, \binits{S.}}:
\batitle{Dimensional study of covid-19 via fractal functions}.
\bjtitle{The European Physical Journal Special Topics}
\bvolume{232}(\bissue{7}),
\bfpage{1061}--\blpage{1070}
(\byear{2023})
\end{barticle}
\endbibitem

%%% 14
\bibitem[\protect\citeauthoryear{Verma and Viswanathan}{2019}]{verma2019revisit}
\begin{barticle}
\bauthor{\bsnm{Verma}, \binits{S.}},
\bauthor{\bsnm{Viswanathan}, \binits{P.}}:
\batitle{A revisit to $\alpha$-fractal function and box dimension of its graph}.
\bjtitle{Fractals}
\bvolume{27}(\bissue{06}),
\bfpage{1950090}
(\byear{2019})
\end{barticle}
\endbibitem

%%% 15
\bibitem[\protect\citeauthoryear{Fern{\'a}ndez-Mart{\'\i}nez and S{\'a}nchez-Granero}{2014}]{fernandez2014fractal}
\begin{barticle}
\bauthor{\bsnm{Fern{\'a}ndez-Mart{\'\i}nez}, \binits{M.}},
\bauthor{\bsnm{S{\'a}nchez-Granero}, \binits{M.A.}}:
\batitle{Fractal dimension for fractal structures: A hausdorff approach revisited}.
\bjtitle{Journal of Mathematical Analysis and Applications}
\bvolume{409}(\bissue{1}),
\bfpage{321}--\blpage{330}
(\byear{2014})
\end{barticle}
\endbibitem

%%% 16
\bibitem[\protect\citeauthoryear{Fraser et~al.}{2018}]{fraser2018assouad}
\begin{barticle}
\bauthor{\bsnm{Fraser}, \binits{J.M.}},
\bauthor{\bsnm{Miao}, \binits{J.J.}},
\bauthor{\bsnm{Troscheit}, \binits{S.}}:
\batitle{The assouad dimension of randomly generated fractals}.
\bjtitle{Ergodic Theory and Dynamical Systems}
\bvolume{38}(\bissue{3}),
\bfpage{982}--\blpage{1011}
(\byear{2018})
\end{barticle}
\endbibitem

%%% 17
\bibitem[\protect\citeauthoryear{Wang and Yu}{2013}]{wang2013fractal}
\begin{barticle}
\bauthor{\bsnm{Wang}, \binits{H.-Y.}},
\bauthor{\bsnm{Yu}, \binits{J.-S.}}:
\batitle{Fractal interpolation functions with variable parameters and their analytical properties}.
\bjtitle{Journal of Approximation Theory}
\bvolume{175},
\bfpage{1}--\blpage{18}
(\byear{2013})
\end{barticle}
\endbibitem

%%% 18
\bibitem[\protect\citeauthoryear{Barnsley et~al.}{1989}]{barnsley1989hidden}
\begin{barticle}
\bauthor{\bsnm{Barnsley}, \binits{M.F.}},
\bauthor{\bsnm{Elton}, \binits{J.}},
\bauthor{\bsnm{Hardin}, \binits{D.}},
\bauthor{\bsnm{Massopust}, \binits{P.}}:
\batitle{Hidden variable fractal interpolation functions}.
\bjtitle{SIAM Journal on Mathematical Analysis}
\bvolume{20}(\bissue{5}),
\bfpage{1218}--\blpage{1242}
(\byear{1989})
\end{barticle}
\endbibitem

%%% 19
\bibitem[\protect\citeauthoryear{Yun}{2019}]{yun2019hidden}
\begin{barticle}
\bauthor{\bsnm{Yun}, \binits{C.-H.}}:
\batitle{Hidden variable recurrent fractal interpolation functions with function contractivity factors}.
\bjtitle{Fractals}
\bvolume{27}(\bissue{07}),
\bfpage{1950113}
(\byear{2019})
\end{barticle}
\endbibitem

%%% 20
\bibitem[\protect\citeauthoryear{Vijender}{2019}]{vijender2019approximation}
\begin{barticle}
\bauthor{\bsnm{Vijender}, \binits{N.}}:
\batitle{Approximation by hidden variable fractal functions: a sequential approach}.
\bjtitle{Results in Mathematics}
\bvolume{74}(\bissue{4}),
\bfpage{192}
(\byear{2019})
\end{barticle}
\endbibitem

%%% 21
\bibitem[\protect\citeauthoryear{Navascu{\'e}s}{2005}]{navascues2005fractal}
\begin{barticle}
\bauthor{\bsnm{Navascu{\'e}s}, \binits{M.A.}}:
\batitle{Fractal polynomial interpolation}.
\bjtitle{Zeitschrift f{\"u}r Analysis und ihre Anwendungen}
\bvolume{24}(\bissue{2}),
\bfpage{401}--\blpage{418}
(\byear{2005})
\end{barticle}
\endbibitem

%%% 22
\bibitem[\protect\citeauthoryear{Navascu{\'e}s}{2007}]{navascues2007non}
\begin{barticle}
\bauthor{\bsnm{Navascu{\'e}s}, \binits{M.}}:
\batitle{Non-smooth polynomials}.
\bjtitle{Int. J. Math. Anal}
\bvolume{1}(\bissue{4}),
\bfpage{159}--\blpage{174}
(\byear{2007})
\end{barticle}
\endbibitem

%%% 23
\bibitem[\protect\citeauthoryear{Akhtar et~al.}{2017}]{akhtar2017box}
\begin{barticle}
\bauthor{\bsnm{Akhtar}, \binits{M.N.}},
\bauthor{\bsnm{Prasad}, \binits{M.G.P.}},
\bauthor{\bsnm{Navascu{\'e}s}, \binits{M.}}:
\batitle{Box dimension of $\alpha$-fractal function with variable scaling factors in subintervals}.
\bjtitle{Chaos, Solitons \& Fractals}
\bvolume{103},
\bfpage{440}--\blpage{449}
(\byear{2017})
\end{barticle}
\endbibitem

%%% 24
\bibitem[\protect\citeauthoryear{Agathiyan et~al.}{2022}]{agathiyan2022construction}
\begin{barticle}
\bauthor{\bsnm{Agathiyan}, \binits{A.}},
\bauthor{\bsnm{Gowrisankar}, \binits{A.}},
\bauthor{\bsnm{Priyanka}, \binits{T.}}:
\batitle{Construction of new fractal interpolation functions through integration method}.
\bjtitle{Results in Mathematics}
\bvolume{77}(\bissue{3}),
\bfpage{122}
(\byear{2022})
\end{barticle}
\endbibitem

\end{thebibliography}
    % \bibstyle{abbrv}
    \bibliographystyle{sn-mathphys-num}

\end{document}